\documentclass[prd,aps,floats,twocolumn,showpacs]{revtex4}
\usepackage[dvips]{epsfig}
\begin{document}
\voffset 15mm
\title{Wang-Landau/Multibondic Cluster Simulations for Second-Order Phase 
       Transitions}

\author{ Bernd A. Berg$^{\rm \,a,b}$ and Wolfhard Janke$^{\rm \,c}$ }

\affiliation{ $^{\rm \,a)}$ Department of Physics, Florida State 
University, Tallahassee, FL 32306, USA\\
$^{\rm \,b)}$ School of Computational Science, 
Florida State University, Tallahassee, FL 32306, USA\\
$^{\rm \,c)}$ Institut f\"ur Theoretische Physik and Centre for 
Theoretical Sciences (NTZ), Universit\"at Leipzig, Augustusplatz 10/11, 
04109 Leipzig, Germany} 

\date{November 15, 2006} 

\begin{abstract}
For a second-order phase transition the critical energy range of interest
is larger than the energy range covered by a canonical Monte Carlo 
simulation at the critical temperature. Such an extended energy range 
can be covered by performing a Wang-Landau recursion for the spectral 
density followed by a multicanonical simulation with fixed weights. But 
in the conventional approach one loses the advantage due to cluster 
algorithms. A cluster version of the  Wang-Landau recursion together 
with a subsequent multibondic simulation improves for 2D and 3D Ising 
models the efficiency of the conventional Wang-Landau/multicanonical 
approach by power laws in the lattice size. 
In our simulations real gains in CPU time reach two orders of magnitude. 
\end{abstract}
\maketitle

Equilibrium properties of statistical physics systems are often 
estimated by Markov chain Monte Carlo (MCMC) simulations \cite{Gu04}. 
In many cases one is interested in calculating expectation values for 
a range of temperatures with respect to the Gibbs canonical ensemble. 
It has turned out that instead of performing simulations of the 
canonical ensemble at distinct temperatures it is often advantageous
to combine them into one simulation of a ``generalized'' ensemble
\cite{ToVa77,BeNe92,LyMa92,WaLa01}; for reviews 
see~\cite{HaOk99,BBook,WJreview}. 

While the power of generalized ensembles is well documented for 
first-order phase transitions and complex systems such as spin glasses 
and peptides (small proteins), this is not the case for second-order
phase transitions, although convenience of such applications is 
claimed by Landau and collaborators \cite{La04}.
However they lose the crucial advantage which cluster algorithms 
\cite{SwWa87,Wo89} provide for MCMC simulations of second-order 
phase transitions. Here we present a generalization to cluster 
algorithms. To keep the paper accessible for non-experts, we restrict 
our investigations to 2D and 3D Ising models, while the points 
made are generally valid for cluster algorithms.

In MCMC simulations of second-order phase transitions one wants
to cover the scaling region in which many physical observables
diverge with increasing lattice size. So we ask the question:
How large is the energy range of this region on a finite lattice 
and is it eventually already covered by a single canonical 
simulations at the (infinite volume) critical temperature $T_c=1/\beta_c$? 

\begin{figure}[tb] \vspace{-2mm} \begin{center}
\epsfig{figure=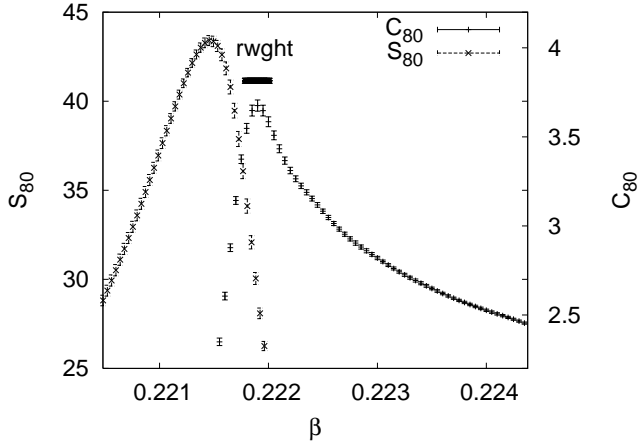,width=\columnwidth} \vspace{-4mm}
\caption{Canonical (indicated by ``rwght'') versus desired
(entire $\beta$ axis) reweighting range on an $80^3$ lattice. 
\label{fig_rwght} }
\end{center} \vspace{-3mm} \end{figure}

For simplicity our lattices are of shape $L^D$ and periodic boundary 
conditions are assumed.  We denote the probability density of the energy
from a canonical MCMC simulation by $P(E)$. Finite-size scaling (FSS) 
arguments \cite{PeVi02} imply $C\sim L^{\alpha/\nu}$ for the 
specific heat at $\beta_c$, where $\alpha$ and $\nu$ are, respectively, 
the critical exponents of the specific heat $C$ and the correlation 
length $\xi$. A second-order phase transition requires $\nu>0$. Let us 
first assume $\alpha>0$. The fluctuation-dissipation theorem gives
\begin{equation} \label{Efluctuation}
 \left\langle (E - \widehat{E})^2\right\rangle \sim L^{D+\alpha/\nu}\
 {\rm where}~~ \widehat{E}= \langle E\rangle\,, 
\end{equation}
implying for the range covered by the simulation at $\beta_c$
\begin{equation} \label{delE}
  \triangle E =
  \left|E_{0.75}-E_{0.25}\right| \sim L^{D/2+\alpha/2\nu}\,,
\end{equation}
where $E_q$, $q=0.25$ and $q=0.75$, are fractiles 
of the energy distribution \cite{BBook}. 
In the vicinity of $\beta_c$ ($A$ constant)
\begin{equation} \label{Ebeta}
  \widehat{E}(\beta)/L^D = \widehat{E}(\beta_c)/L^D
                  + A\,(\beta-\beta_c)^{1-\alpha}\,,
\end{equation}
and using the hyperscaling relation \cite{PeVi02} $\alpha=2-D\nu$, 
Eq.~(\ref{delE}) translates into a reweighting range
\begin{equation} \label{dbcano}
  \triangle \beta \sim L^{-1/\nu}\ .
\end{equation}
The desired reweighting range is determined by the 
need to cover the maxima of all divergent observables measured. Let
the maximum value of such an observable $\widehat{S}_L(\beta)$ be
$\widehat{S}_L^{\max}=\widehat{S}_L(\beta_L^{\max})$ and denote 
the critical exponent of $S$ by $\sigma$. Then FSS theory implies 
\begin{equation} \label{Smax}
  \widehat{S}_L^{\max} \sim L^{\sigma/\nu}\ .
\end{equation}
Reweighting has to cover a reasonable range about the maximum, say 
from $\beta_L^{r-}$ to $\beta_L^{r+}>\beta_L^{r-}$ defined as 
solutions of 
\begin{equation} \label{Srange}
  \widehat{S}_L(\beta) = r\, \widehat{S}_L^{\max},~~0<r<1\,,
\end{equation}
which becomes large for $r$ small. 
We define $\beta_L^r\in\{\beta_L^{r-}, \beta_L^{r+}\}$ to be the 
$\beta_L^{r^{\pm}}$ value which is further away from $\beta_c$ 
than the other and assume 
\begin{equation} \label{delbr}
  \triangle \beta_L^r = \left|\beta_L^r-\beta_c\right| =
  a^r\,L^{-\kappa}\,,
\end{equation}
where $a^r$ and $\kappa>0$ are constants ($\kappa$ independent of $r$).
For sufficiently large $L$ we suppose that
\begin{equation} \label{S_Lbr}
  \widehat{S}_L(\beta_L^r) = S^{\rm reg} + A\,\left(\triangle
  \beta_L^r\right)^{-\sigma} 
\end{equation}
holds, where $S^{\rm reg}$ is a regular background term. Combining
Eqs.~(\ref{Smax}), (\ref{delbr}), and (\ref{S_Lbr}) we conclude
\begin{equation} \label{kappa}
  \kappa = 1/\nu\,,
\end{equation}
i.e., the desired range (\ref{delbr}) scales in the same way as the
canonical range (\ref{dbcano}). However, the proportionality factor
$a^r$ can be much larger than the one encountered for the canonical
range. With the modest value $r=2/3$ this point is made in 
Fig.~\ref{fig_rwght} for the 3D Ising model on an $80^3$ lattice.
We plot the specific heat $C(\beta)$ and for $S(\beta)$ the first 
structure factor (see, e.g., Ref.~\cite{Stanley}), whose maximum 
scales $\sim L^{\gamma/\nu}$. The desired reweighting range is more 
than 17 times larger than the canonical reweighting range from a 
simulation at $\beta_L^{\max}$ of the specific heat (in realistic 
applications one does not know $\beta_c$ a-priori and $\beta_L^{\max}$ 
of a suitable observable is a good substitute).

Using the same line of arguments for a logarithmic singularity 
\begin{equation} \label{S_ln}
  S(\beta) = S^{\rm reg} - A\,\ln|\beta-\beta_c|
\end{equation}
one finds that the exponent $\kappa$ in Eq.~(\ref{delbr}) is no longer 
independent of $r$, but 
\begin{equation} \label{kappa_ln}
  \kappa\ =\ r/\nu\ .
\end{equation}
While the canonical reweighting range scales still $\sim L^{-1/\nu}$,
the desired reweighting range becomes $\sim L^{r/\nu}$, so that the
ratio desired/canonical diverges $\sim L^{(1-r)/\nu}$. With $S=C$ the 
2D Ising model provides an example.

In conclusion many more than one canonical simulation are typically
needed to cover a relevant part of the scaling region of a second-order
phase transition. In principle this can be done by patching canonical 
simulations from several temperatures together, relying on a 
multi-histogram approach \cite{FeSw89}. Besides that dealing with many 
simulations is tedious, weaknesses of these approaches are that the 
histograms fluctuate independently and that their patching has to be 
done in regions where the statistics is reduced due to the decline of 
the number of histograms entries. More stable estimates are obtained 
by constructing a generalized ensemble, which allows the random walker 
to cover the entire region of interest. This requires two steps:

\begin{enumerate}

\item Obtain a working estimate of the weight factors.

\item Perform a MCMC simulation with fixed weights.

\end{enumerate}

To be definite we confine our discussion to the multicanonical (MUCA)
simulations~\cite{BeNe92}. Extension to cluster algorithms are known 
\cite{JaKa95,Lo06}. We will rely on multibondic (MUBO) simulations
\cite{JaKa95}. This defines step~2, but leaves still many options
open to deal with step~1.
``Working estimate'' means that the approximation of the weights of 
the generalized ensemble is good enough so that the energy range in 
question is covered in step~2. Historically step~1 has been one
of the stumbling blocks of umbrella sampling: 
``The difficulty of finding such weighting factors has prevented wide 
applications of the umbrella sampling method to many physical systems"
\cite{LiSc88}. 
Most convenient is to have an efficient 
general purpose recursion for step~1 at hand. While designs were 
reported in a number of papers \cite{recursions}, see also 
Refs.~\cite{BBook,WJreview,Lo06}, the winning approach appears to be the one of 
Wang and Landau (WL)~\cite{WaLa01} (although somewhat surprisingly we 
found only one comparative study \cite{Ok03}). 

The WL recursion differs fundamentally from the earlier approaches 
by iterating the weight at energy $E$ {\em multiplicatively\/} with a factor
$f_{\rm WL}>1$ rather than additively. At a first glance the WL approach 
is counter-intuitive, because the correct iteration of the weight 
factor close to the desired fixed point is obviously proportional 
to one over the number of histogram entries $H(E)$ and not to
$1/f_{\rm WL}^{H(E)}$. However, what matters is a rapid approach to 
a working estimate. The advantage of the WL over the other recursions 
is that it moves right away rapidly through the targeted energy range. 
When it is sufficiently covered, the iteration factor is refined by
$f_{\rm WL}\to\sqrt{f_{\rm WL}}$, so that it approaches 1 rapidly. Once 
the system cycles with frozen weights through the desired energy range 
the goal of a working estimate has been reached and the WL recursion 
is no longer needed \cite{comment}. We now generalize this approach 
to cluster algorithms.

We use the energy function of the $q$-state Potts models, 
\begin{equation} \label{E}
  E = - 2 \sum_{\langle ij\rangle} \delta_{q_iq_j}\ ,
\end{equation}
where the sum is over the nearest-neighbor sites of a $D$-dimensional 
cubic lattice of $N=L^D$ Potts spins, which take the values $q_i=1,
\dots,q$. The factor of two has been introduced so that $q=2$ 
matches with the energy and $\beta$ conventions of the Ising model 
literature.

In the Fortuin-Kasteleyn (FK) cluster language \cite{FK72} the Potts
model partition is written as
\begin{eqnarray} \nonumber
  Z_{\rm FK} &=&\sum_{\{q_i\}} \sum_{\{b_{ij}\}} Z(\{q_i\},\{b_{ij}\})
  ~~~{\rm with}\\ \label{Z_FK}
  Z(\{q_i\},\{b_{ij}\}) &=& \prod_{\langle ij\rangle} 
  \left[ a\,\delta_{q_iq_j}\,\delta_{b_{ij} 1} 
  + \delta_{b_{ij} 0} \right]
\end{eqnarray}
where $a=e^{2\beta}-1$. For a fixed configuration $\{q_i\}$ of Potts 
states the Swendsen-Wang updating procedure \cite{SwWa87} is to 
generate bonds variables $b_{ij}$ (simply called bonds in the 
following) on links with $\delta_{q_iq_j}=1$: Bonds $b_{ij}=1$ are
generated with probability $p$ and bonds $b_{ij}=0$ with probability 
$q$ so that $p/q=a$ and $p+q=1$ holds. This gives
$p=1-e^{-2\beta}$ for $b_{ij}=1$ and $q=1-p=e^{-2\beta}$ for $b_{ij}=0$.
On $\delta_{q_iq_j}=0$ links we have $b'_{ij}=0$ with probability one. 
We call bonds with $b_{ij}=1$ active or set. A cluster of spins is 
defined as a set of spins connected by active bonds and an update
is to flip entire clusters of spins, $\{q_i\}\to \{q'_i\}$. 

Let us denote the number of active bonds by $B$. The MUBO partition 
function \cite{JaKa95} is defined by
\begin{equation} \label{Z_MUBO}
   Z_{\rm MUBO}=\sum_{\{q_i\}} \sum_{\{b_{ij}\}} Z(\{q_i\},\{b_{ij}\})
   \,W(B)
\end{equation}
where a bond weight factor $W(B)$ has been introduced. A valid updating 
procedure for the configurations of this partition function is formulated 
in the following.

A. For $q_i\ne q_j$ a bond is never set. This applies to the initial as 
well as to the updated bond on this link, so that $B$ does not change.
B.~For $q_i=q_j$ there are two possibilities:

\begin{enumerate}

\item The initial bond is not set, $b_{ij}=0$. Then $B'=B$ for 
      $b'_{ij}=0$ and $B'=B+1$ for $b'_{ij}=1$. The updating
      probabilities are
\begin{eqnarray} \label{P10}
  P_1(0\to 0) &=& \frac{q\,W(B)}{q\,W(B)+p\,W(B+1)}
\end{eqnarray}
and $P_1(0\to 1)=1-P_1(0\to 0)$.

\item The initial bond is set, $b_{ij}=1$. Then $B'=B-1$ for 
      $b'_{ij}=0$ and $B'=B$ for $b'_{ij}=1$. The updating
      probabilities are
\begin{eqnarray} \label{P20}
  P_2(1\to 0) &=& \frac{q\,W(B-1)}{q\,W(B-1)+p\,W(B)}
\end{eqnarray}
and $P_2(1\to 1)=1-P_2(1\to 0)$.

\end{enumerate}

After the configuration is partitioned into clusters \cite{cluster}, 
the update 
is completed by assigning with 
uniform probability a spin in the range $1,\dots,q$ to each cluster.

In its generalization to cluster algorithms the WL recursion 
updates then $\ln W(B)$ according to
\begin{equation} \label{WLcl} 
  \ln W(B)\ \to\ \ln W(B) - a_{\rm WL}\,,\ \
  a_{\rm WL}=\ln(f_{\rm WL})\,,
\end{equation}
whenever a configuration with $B$ bonds is visited. All recursions are
started with $a_{\rm WL}=1$ and we iterate $a_{\rm WL}\to a_{\rm WL}/2$ 
according to the following criteria:

\begin{enumerate}

\item The Markov chain just cycled from $\overline{B}_L^{r-}$ to 
$\overline{B}_L^{r+}$ and back. Here $\overline{B}_L^{r-}$ and 
$\overline{B}_L^{r+}$ are bond estimates corresponding to
$\beta_L^{r-}$ and $\beta_L^{r+}$, respectively, determined by 
short canonical simulations.

\item The bond histogram $h(B)$, measured since the last iteration,
fulfilled a flatness criterion $h_{\min}/h_{\max} > cut$, where $cut$ 
was equal to $1/3$ in most of our runs.

\item We freeze the weights after a last iteration is performed with 
      the desired minimum value $a_{\rm WL}^{\min}$.

\end{enumerate}

\begin{table}[tb]
\caption{3D Ising model simulations on $L^3$ lattices. 
\label{tab_I3DMUBO}} \smallskip 
\centering
\begin{tabular}{cccccr}  \hline\hline
$L$&$\beta_L^{r-}$&$\beta_L^{r+}$&$a_{\rm WL}^{\min}$&recursion&production 
\\ \hline
20&0.210\,649&0.233\,690&$2^{-18}$&$~19\,962$ &$32\times  32\,768$\\
30&0.216\,443&0.229\,336&$2^{-18}$&$~27\,344$ &$32\times  32\,768$\\
44&0.218\,545&0.227\,013&$2^{-19}$&$~33\,266$ &$32\times  65\,536$\\
56&0.219\,755&0.225\,914&$2^{-19}$&$~56\,323$ &$32\times  65\,536$\\
66&0.220\,063&0.224\,709&$2^{-21}$&$~62\,884$ &$32\times 131\,072$\\
80&0.220\,482&0.224\,377&$2^{-21}$&$108\,618$ &$36\times 131\,072$\\ \hline\hline
\end{tabular} \end{table} \vspace*{0.2cm}

After a short equilibration run, measurements are performed during the 
subsequent simulation with fixed weights, each tuned to approximately 
1$\,$000 cycling events. Canonical expectation values at inverse 
temperature $\beta$, $\beta_L^{r-}\le\beta\le \beta_L^{r+}$ are 
obtained by reweighting (\ref{Z_MUBO}). Table~\ref{tab_I3DMUBO} 
gives an overview of our 3D Ising model statistics. The effectiveness 
of the recursion is seen from the fact that it takes never more than 
3\% percent of the statistics used for production (these numbers are 
in sweeps). Similarly the initial simulations, which determine 
$\overline{B}_L^{r-}$ and $\overline{B}_L^{r+}$, take less than 3\%.

\begin{figure}[tb] \vspace{-2mm} \begin{center}
\epsfig{figure=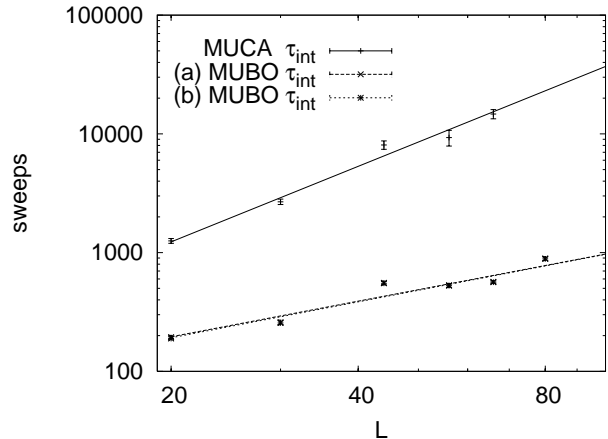,width=\columnwidth} \vspace{-4mm}
\caption{$\tau_{\rm int}(L)$ for the 3D Ising model (see text).
\label{fig_tauI3D}}
\end{center} \vspace{-3mm} \end{figure}

From the production statistics we calculate integrated autocorrelation 
times $\tau_{\rm int}$ 
and compare them in Fig.~\ref{fig_tauI3D} with those of a MUCA 
simulation. From the MUBO time series we calculated $\tau_{\rm int}$ 
for (a)~energies and (b)~bonds and found the results almost identical 
(slightly higher for the energies, but indistinguishable on the scale 
of the figure).  For MUCA the estimates are from energies. Up to a 
constant factor practically identical results are obtained from cycling 
times. In our code one MUCA sweep was about three times faster than
one MUBO sweep. 

The critical slowing down is $\sim L^z$. For the dynamical critical 
exponent we find $z=2.22\,(11)$ for MUCA and $z=1.05\,(5)$ for MUBO.
So the performance gain is a bit better than linear with the lattice 
size $L$.  The data in Fig.~\ref{fig_tauI3D} scatter more than one 
might have expected about the fits because our $\beta_L^{r-}$ and 
$\beta_L^{r+}$ values are based on MCMC estimates, which are by 
themselves noisy. Our exponent for cluster updating is significantly 
higher than the one estimated from simulations at $\beta_c$, $z=0.50\, 
(3)$, for the Swendsen-Wang algorithm \cite{Wolff89}. The reason is 
that the efficiency of the cluster algorithm deteriorates off the 
critical point, even when one is still in the scaling region. 
Therefore, we think that our exponent of $z\approx 1$ reflects the 
slowing down in real application more accurately  than the small value 
of the literature. In particular the cluster algorithm becomes rather 
inefficient for calculating the long tail of the specific heat for 
$\beta > \beta_L^{\max}$.

\begin{figure}[t] \vspace{-2mm} \begin{center}
\epsfig{figure=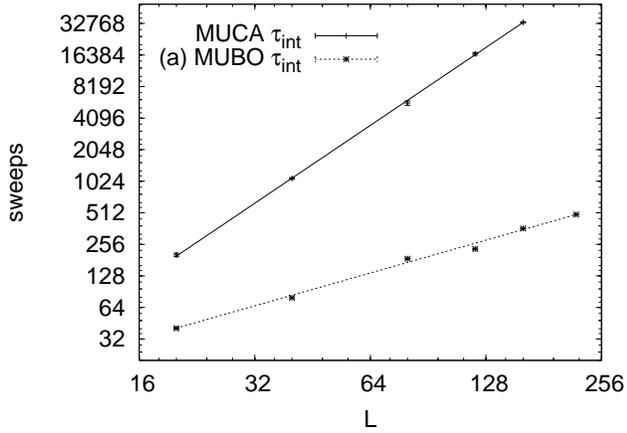,width=\columnwidth} \vspace{-4mm}
\caption{$\tau_{\rm int}(L)$ for the 2D Ising model.\label{fig_tauI2D}}
\end{center} \vspace{-3mm} \end{figure}

In Fig.~\ref{fig_tauI2D} we show integrated autocorrelation times from 
simulations of the 2D Ising model for which we adjusted our simulation 
parameter to cover the full width at half-maximum of the specific heat. 
This corresponds to $r=1/2$ in Eq.~(\ref{kappa_ln}). The dynamical 
critical exponent takes than the values $z=2.50\,(4)$ for MUCA and 
$z=1.04\,(2)$ for MUBO. The MUCA value reflects that the number of 
canonical simulations needed to cover the desired energy range grows 
now $\sim L^{1/2}$, while the canonical critical value is approximately 
two \cite{BBook,LaBi}.

Finally we remark that the efficiency of simulations of second-order
phase transitions can presumably be further improved by optimizing
the weights with respect to cycling along the lines introduced in
Ref.~\cite{THT04}.

\acknowledgments
This work started while BB was the Leibniz Professor at Leipzig
University. In part it was supported by the US Department of Energy 
under contract DE-FG02-97ER41022 and by the Deutsche 
Forschungsgemeinschaft (DFG) under contract JA 483/23-1. 
After submitting this paper we learned from Professor Y. Okabe that 
Eq.~(\ref{WLcl}) was previously derived in Ref.~\cite{YaKa02}.


\begin{thebibliography}{19}
\bibitem{cluster} This is the Swendsen-Wang approach \cite{SwWa87}. 
                  To work out the Wolff version \cite{Wo89} is left 
                  as an exercise to the reader.
\end{thebibliography}
\end{document}